# Superior performance of multilayered fluoropolymer films in low voltage electrowetting

*Dimitrios P. Papageorgiou[1], Angeliki Tserepi[2], Andreas G. Boudouvis[1], Athanasios G. Papathanasiou[1,*]*

[1]School of Chemical Engineering, National Technical University of Athens, GR-15780, Athens, Greece

[2]Institute of Microelectronics, NCSR "Demokritos", GR-15310, Aghia Paraskevi, Athens, Greece

Author email address:

 dpapag@chemeng.ntua.gr (D. P. P.), atserepi@imel.demokritos.gr (A. T.), boudouvi@chemeng.ntua.gr (A. G. B.)

[*]Author to whom correspondence should be addressed.

School of Chemical Engineering, Dept. II
National Technical University of Athens
Zografou Campus, 15780
Athens,
GREECE
Office No: H1.502
Tel.:+30-2107723290;
Fax: +30-2107723155;
e-mail: pathan@chemeng.ntua.gr;
web: www.chemeng.ntua.gr/people/pathan

**Abstract**

The requirement for low operational voltage in electrowetting devices, met using thin dielectrics, is usually connected with serious material failure issues. Dielectric breakdown (visible as electrolysis) is frequently evident slightly beyond the onset of the contact angle saturation. Here, plasma enhanced chemical vapor deposition (PECVD) is used to deposit thin fluorocarbon films prior to the spin-coating of Teflon® amorphous fluoropolymer on tetraethoxysilane (TEOS) substrates. The resulting multilayered hydrophobic top coating improves the electrowetting performance of the stack, by showing high resistance to dielectric breakdown at high applied voltages and for continuous long term application of DC and AC voltage. Leakage current measurements during electrowetting experiments with the proposed composite coating showed that current remains fairly constant at consecutive electrowetting tests in contrast to plain Teflon® coating in which material degradation is evident by a progressive increase of the leakage current after multiple electrowetting tests. Since the proposed composite coating demonstrates increased resistance to material failure and to dielectric breakdown even at thin configurations, its integration in electrowetting devices may impact their reliability, robustness and lifetime.

I. **INTRODUCTION**

Electrowetting on dielectric (EW) is used to modify the wettability of a dielectric solid substrate by a conductive liquid, through a suitable application of an external voltage. A voltage applied between a conductive drop sitting on a flat electrode coated by the dielectric induces charge accumulation on the liquid/dielectric solid interface. The accumulated electrostatic energy reduces the liquid/solid interfacial energy enhancing the wetting of the solid by the liquid, manifested as reduction of the drop contact angle. In an oil ambient, EW can provide more than 100° of contact angle modulation reversibly, with high fast response to actuation[1] and as a result EW is utilised in various lab-on-chip devices[2], variable focus lenses[3] and electronic displays[4]. The dependence of the contact angle, $\theta_V$, on the applied voltage, $V$, is given by the Lippmann[5] equation, which reads:

$$\cos\theta_V = \cos\theta_Y + \frac{\varepsilon_0\varepsilon_r}{2d\gamma}V^2, \qquad (1)$$

where $\theta_Y$ is the Young's contact angle, $\gamma$ is the liquid surface tension, $d$ is the thickness of the dielectric with dielectric constant $\varepsilon_r$, and $\varepsilon_0$ is the permittivity of vacuum. Experiments show that beyond a critical voltage, namely $V_s$, the contact angle (CA) reaches a lower limit, in contradiction to eq.(1) which predicts complete wetting i.e. $\theta_V = 0°$ at sufficiently high applied voltage. This effect is widely known as CA saturation that limits the EW response to the applied voltage. Recent studies attribute CA saturation to dielectric breakdown[6], dielectric charge trapping[7, 8] and air ionization[5].

Device operation preferably requires low voltages to take advantage of the maximum CA modulation with minimum energy requirement. According to the Lippmann equation (cf. eq. 1) low applied voltages could be attained either by reducing the interfacial tension between the liquid and oil phases ($\gamma$), i.e. the use of ionic surfactants[9] or by decreasing the overall dielectric thickness. On the one hand, the use of ionic species can induce the effect of electrolysis of an EW system at dielectric defects[10] and, on the other hand, reduced dielectric thickness increases the chance of failure due to film defects. Therefore, in low voltage EW systems these matters have to be carefully considered in the resulting EW device design.

Usually in low voltage EW applications a thin multilayer dielectric stack is used, comprising a main insulating layer [i.e. $SiO_2$, $Si_3N_4$, tetraethoxysilane (TEOS) etc.] and a hydrophobic top coating (i.e. Teflon®, Cytop® amorphous fluoropolymers etc.)[11]. Amorphous fluoropolymers (aFP) are inherently more porous than the insulating layer and feature poor adhesion to oxide substrates. High porosity is usually related to charge injection through the overall hydrophobic dielectric, which could ultimately lead to dielectric breakdown[10]. Moreover, the adhesion of the hydrophobic top coating to the main dielectric, which significantly affects the overall mechanical stability of the stack, is a crucial matter as the deposited films fail to endure persistent EW testing[12]. EW systems with protection against dielectric breakdown have also been reported, utilizing recovering from electrolysis due to anodization[13]. Furthermore, reliable and low-voltage EW is shown on thin Parylene films through continuous DC and AC electrowetting tests[14].

In this work, we use plasma-enhanced chemical vapor deposition to deposit thin fluorocarbon (FC) films on tetraethoxysilane (TEOS) substrates, for the development of hydrophobic dielectrics. The use of the plasma-deposited FC films is advantageous due to conformal deposition, thickness uniformity and better adhesion to oxide substrates[15, 16]. However, stand-alone plasma FC films exhibit large water contact angle hysteresis (CAH) due to increased surface roughness compared to spin-coated aFP, as well as surface ageing of the fluorinated films. For that reason, a spin-coated Teflon® layer is followed on top of the plasma-deposited FC. The proposed composite hydrophobic top coating utilizes the adhesion benefits of plasma FC on oxide substrates and the surface smoothness of the spin-coated commercial Teflon® film.

Reported herein is a composite hydrophobic top coating tested at high applied voltages (even beyond the critical saturation voltage) for a number of repeatable EW tests. The composite coating showed resistance to dielectric breakdown, high CA modulation range (>110°), and reversibility in multiple EW experiments. Measurements of the leakage current density through the composite coating/TEOS assembly proved to be repeatable in subsequent EW tests compared to Teflon® coated samples. Therefore, device implementation of the proposed composite coating could lead to more efficient and reliable EW devices.

## II. EXPERIMENTAL SETUP AND MEASUREMENTS

Hydrophobic dielectric stacks are fabricated for conducting EW experiments on (resistivity, 1-10 Ω/cm). The hydrophobic dielectric stacks consist of a main dielectric, developed on phosphorus-doped Si wafers, and a hydrophobic top coating. TEOS is used as the main insulating layer and Teflon® as hydrophobic top coating. Special attention is given on the preparation of the top coating. In addition to the standard technique of improving the adhesion of spin-coated Teflon® on top of oxides, we used and tested an intermediate layer of plasma-deposited FC. The samples tested are compared in terms of breakdown prevention, resistance to material failure, and leakage current, which are critical aspects in the operation of EW devices.

The standard samples are prepared using fluorosilanes that are widely used to improve the adhesion of Teflon® AF 1600 to untreated Si and glass substrates[17]. Perfluorooctyltriethoxysilane solution is spin-coated onto TEOS and the coated wafers are then heated at 95 °C for 15 min. Teflon® is diluted at Fluorinert® Fluid FC-77 solvent; and then spun on top of the fluorosilane layer. After the spinning process, the sample is baked in air, at 95 °C for 5 min.

In this work, plasma-deposited FC is suggested as an adhesion promoter layer for the spin-coated Teflon®[18]. In particular, thin plasma FC films (10-60 nm) are formed by plasma enhanced chemical vapor deposition (PECVD) using inductively coupled plasma (ICP)[19]. Teflon® (30-100 nm) is then spin-coated on the FC layer.

Throughout this article, the type of hydrophobic top coating that consists of Teflon® and fluorosilane primer will be referred as "*Teflon® coating*", whereas the type of coating that consists of a thin plasma-deposited FC film and Teflon® will be referred as "*composite coating*".

Verification of the thicknesses of the oxide and the top coating layers is performed with a spectroscopic ellipsometer, model M2000 J.A. Woolam Co. (accuracy in the measured thickness ± 0.5 nm). The fabrication of the samples tested is conducted in clean room environment.

EW experiments are performed using water or sodium dodecyl sulfate (SDS) sessile droplets (0.1wt.% SDS in 0.1M NaCl) with conductivities of 160-200 μS/cm and ~11.22 mS/cm respectively. The insulating ambient phase is air or dodecane oil. Both AC (2.3 kHz sine wave) and DC electrowetting is performed.

Static CA measurements are performed through observation of the drop shape (and its reflection) with respect to the sample surface. Contact angle hysteresis measurements are performed by increasing (advancing CA) or decreasing (receding CA) the droplet volume; the difference of the advancing and receding CA values determines CAH.

Measurements of the CA dependence on applied voltage are performed in an in-house built EW setup, described in Papathanasiou et al.[20]. In this setup, the samples are immersed in a completely transparent poly(methyl methacrylate) (PMMA) oblong tank filled with 99+% pure dodecane. Real time image processing software, that is developed in house, is used to analyze the drop shape. EW experiments are also accompanied by leakage current measurements in order to check the insulating properties of the dielectrics tested. Pure insulating performance is usually an indication of pure EW performance.

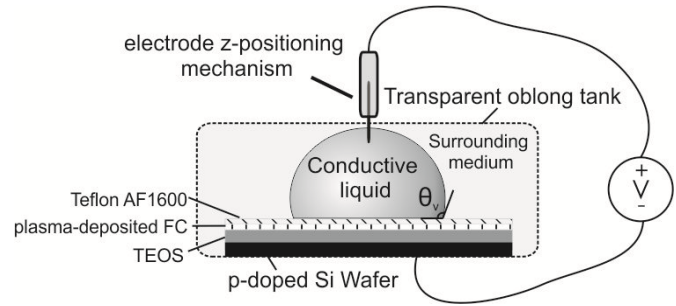

**Figure 1.** Sketch of the experimental setup: The ground electrode is a p-doped Si wafer with TEOS followed by plasma-deposited FC and a Teflon® coating. The sample is preferably submerged in a transparent oblong tank which is filled with the insulating liquid (dodecane). (The layer thicknesses are not shown in scale)

### III. RESULTS AND DISCUSSION

**A. Electrolysis effect inhibition on composite-coated samples**

In most cases, in the saturation regime (when the applied voltage exceeds the critical saturation voltage, $V_s$), bubble generation (i.e. electrolysis) indicates sample deficiency, which limits its functionality for performing further experiments[20]. Here we found that, the integration of plasma FC in the hydrophobic dielectric stack significantly inhibits the electrolysis even at applied voltages considerably higher than the saturation voltage. Plasma FC is deposited on top of the main insulating layer, followed by a spin-coated Teflon® layer to provide the required hydrophobicity.

Our previous findings suggest that one of the factors that could lead to sample failure is possibly poor adhesion between the hydrophobic top coating and the oxide substrate[12]. Nanoscratch testing revealed improved mechanical interlayer properties of the composite coating/TEOS compared to the Teflon® coating/TEOS assembly, confirming the observed improved robustness in EW tests. In that sense, the samples tested herein feature Teflon® and composite coating, and their EW performance is compared in terms of breakdown prevention, EW hysteresis, and resistance to material failure when subjected to multiple tests at voltages beyond the saturation limit. Each AC electrowetting cycle is performed in air and oil ambient as follows: The applied voltage is increased from 0 V (voltage ramp rate is 1 V/s) up to $2.5V_s$. For applied voltages up to $V_s$, as expected, our experimental data are in close agreement with the Lippmann predictions, for all tested samples.

In Figure 2, 300 nm TEOS samples featuring composite and Teflon® coatings are compared in terms of EW performance. For an applied voltage of 18 V, the CA modulation is 110° (from 160° to 50°) for both samples. The composite coating sample exhibits a CA of 30° at 45 V and a consequent maximum CA modulation of 128°. Similar CA modulation (from 150° down to 45°) was also observed by Paneru et al.[21] for both DC and AC electrowetting. In that work, the hydrophobic dielectric comprised only Teflon® film on top of an ITO electrode. However, the Teflon® film thickness was about 2.3 μm, i.e. relatively thick, which limits its applicability in devices due to high required voltages (~100 V). Since the design of EW devices should meet the requirement for low applied voltage coupled with high CA modulation range, thin dielectrics are used in order to achieve operational voltages lower than 20 V.

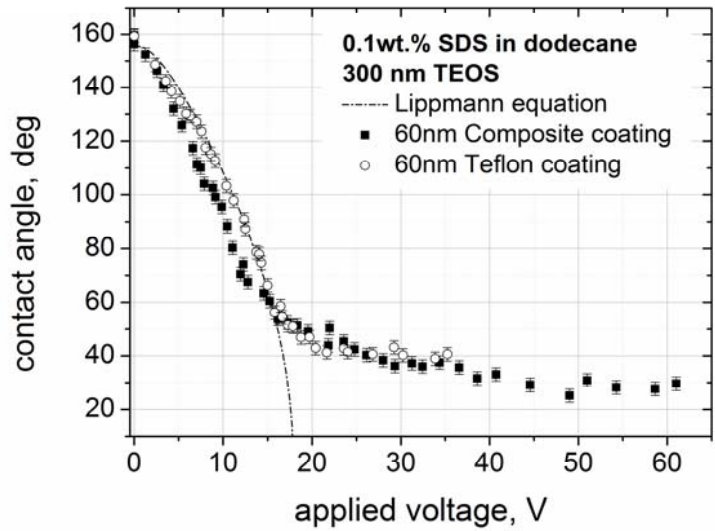

**Figure 2.** Dependence of the CA on the AC applied voltage for the composite and Teflon® coatings respectively.

In the case of Teflon® coating, when the applied voltage reaches approximately $1.4V_s$, bubbles start to rise inside the liquid drop, an indication of electrolysis that limits the sample functionality of performing further EW cycles. Our experiments showed that the robustness of the composite coating in terms of endurance is superior, since it is successfully tested at high voltages up to $2.5V_s$, for multiple EW cycles. Failure indication (i.e. electrolysis) is not noticeable for at least up to 30 EW cycles, however the range of CA modulation gradually decreases from a maximum of 128° to 110°. This gradual loss of performance is evident up to the fourth EW cycle and thereafter CA modulation range remains unaltered, which could be possibly attributed to fluoropolymer charging[8], or similarly, inadequate insulation of the oil ambient[22]. The following Table 1 correlates the applied critical voltage $V_s$, to CA in the saturation regime for the different hydrophobic coatings. The last two columns of Table 1 show the minimum CA ($\theta_{min}$) achieved with respect to the applied voltage.

|              |         |           | AC Electrowetting, 0.1wt.% SDS |           |                  |         |
|              |         |           | In dodecane                    |           |                  |         |
| TEOS Thickness (nm) | Type of Coating | Coating Thickness (nm) | $\theta_s$ (deg) | $V_s$ (Volts) | $\theta_{min}$ (deg) | V (Volts) |
|---|---|---|---|---|---|---|
| 150 | Composite | 55 | 49±1.5 | 13.8±1 | 34±2 | 40±1.5 |
| 300 | composite | 60 | 50±1.5 | 17.3±1 | 30±2 | 45±1.5 |
| 300 | Teflon® | 60 | 50±1.5 | 17.8±1 | 39±2 | 35±1.5 |
| 830 | composite | 200 | 46±1.5 | 32.1±1 | 26±2 | 77±1.5 |

**Table 1.** Specifications of the samples tested.

The tested samples showed very low CAH (CAH ~12° in air ambient) and fairly good EW hysteresis (CA at zero voltage for two subsequent cycles). The EW hysteresis is of the order of ~6° in dodecane ambient. We should stress that even after exceeding twice the critical saturation voltage, namely at 2.5$V_s$, EW hysteresis remains low (≤6°). In all samples, the CA range of modulation is reduced by about ~10° after the first four cycles. Detailed data of CAH and EW hysteresis for each sample are listed in Table 2.

|              |         |           | AC Electrowetting |                |                |
|              |         |           | Water Droplet     |                | 0.1wt.% SDS    |
|              |         |           | In air            | In dodecane    | In dodecane    |
| TEOS Thickness (nm) | Type of Coating | Coating Thickness (nm) | CAH (deg) | EW Hysteresis (deg) | EW Hysteresis (deg) |
|---|---|---|---|---|---|
| 150 | composite | 55 | 12±2.5 | 6±2.5 | 6±2.5 |
| 300 | composite | 60 | 12±2.5 | 6±2.5 | 5±2.5 |
| 300 | Teflon® | 60 | 11±2.5 | 5±2.5 | 5±2.5 |
| 830 | composite | 200 | 12±2.5 | 6±2.5 | 4±2.5 |

**Table 2.** Contact angle hysteresis and electrowetting hysteresis data of the samples tested.

Electrowetting hysteresis, which is mostly attributed to charge trapping at the hydrophobic dielectric, is observed at all samples tested. Figure 3 presents a hysteresis loop for a composite coated sample. In a usual hysteresis loop, at certain applied voltage, the CA is higher when voltage increases (advancing CA) than when voltage decreases (receding CA). However, in plasma FC containing top coatings, this trend is reversed (i.e. CA is higher when voltage decreases) and it is pronounced at voltages close to the saturation onset and higher where leakage or trapping of charge is considerable. This strange behavior was also observed by Bayiati et al.[23], for various plasma FC film thicknesses and for both DC polarities. We suspect that when voltage decreases charge detrapping is slow with respect to voltage ramp rate (1 V/sec), resulting to increased remaining trapped charge which degrades the EW performance and leads to higher CA when voltage decreases. The mechanism related to charge trapping/detrapping phenomena is still under investigation as it strongly depends on the dielectric materials and surfactants used in EW systems.

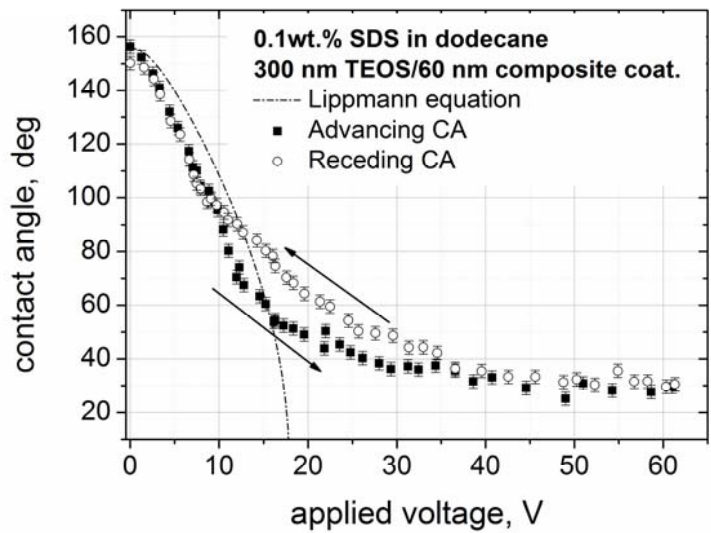

**Figure 3.** EW hysteresis in AC applied voltage of 300 nm TEOS/60 nm composite coating stack.

In that sense, ionic surfactants can reduce the interfacial tension between water and oil surrounding phase and therefore low voltage EW can be achieved. Moreover, in lab-on-chip EW applications ionic content (i.e. SDS etc.) is preferable since ionic content provides favourable conditions for the stability of biological content like proteins, cells or DNA. However, surfactant ions can induce electrolysis at dielectric defects, especially at positive DC applied voltage[10]. It is remarkable that our proposed composite coating shows no sign of electrolysis in repeatable EW tests up to $2.5V_s$, even with SDS solution. Dielectric failure, alternatively, can be prevented by the use of large ions (carrying long alkane chain i.e. SOS, DTAC, DTA-OS etc.) in EW liquids since charge propagation through the dielectric is limited[10]. In that sense, it is possible that the use of the composite coating and the careful selection of the liquid ionic content in applications could further enhance device reliability and robustness.

**B. Leakage Current measurements on Teflon®, plasma-deposited FC, and composite coatings**

Throughout this section, DC electrowetting is performed for leakage current measurements, of a water sessile droplet in air ambient. TEOS (300nm thick) is used as the main insulating layer. Here we test three different fluorocarbon-based top coatings, in particular, spin-coated Teflon®, plain plasma-deposited FC and the proposed composite coating.

In each measurement, the applied voltage is increased in steps of 5 V (the voltage is never turned off between the voltage steps, i.e. constant voltage measurement). This technique is also applied by Gischia et al.[24] to estimate the trapped charge density in planar capacitors featuring porous low-k dielectric materials.

Figure 4 presents the current response to the voltage steps for the tested top coatings. Note that the dielectric constants of plasma-deposited FC and Teflon® are fairly identical so the capacitance per unit area is practically the same for equally thick top coatings. The current spike, evident in the Teflon® coating (Fig. 4a), corresponding to the onset of every voltage step, could be possibly attributed to the charging phase of the capacitor formed by the EW system[20] (see pointing arrows in Fig. 4a). When the voltage is kept constant, the potential traps, existing in the inherently porous Teflon® coating, are gradually filled with charges and consequently the current decreases. As a result, the following phase of the slow and fluctuating current decrease possibly suggests charge trapping process in the hydrophobic dielectric[24]. Maybe structural imperfections such as voids (due to porous structure) induce trapping/de-trapping mechanisms resulting in the observed fluctuating current response.

However, the referred current spike and the following phase of the current decrease is not evident at the composite coating as Fig. 4b shows. The current response to the voltage steps is a quite uniform step function and for applied voltages between 0 - 45 V the current reaches almost abruptly its equilibrium value. Since the composite coating also utilizes spin-coated Teflon® as the upper layer, the charge traps are supposed to appear mostly on the interface between the oxide layer and the hydrophobic coating in the case of the single layered spin-coated Teflon®. The improved adhesion strength between the TEOS and the plasma-deposited fluorocarbon (composite coating) is assumed to reduce or prevent creation of voids at the interface, thus preventing significant charge trapping, more favored in the case of the Teflon® coating.

The current response of the plasma FC (Fig. 4c) is found to be similar with the one of the composite coating sample independently of the plasma FC thickness (15, 30 and 60nm thick plasma FC films were deposited on TEOS). However, the mean value of the leakage current for each voltage step increases significantly if thinner plasma FC layers are used, showing the effect of the thickness of the plasma FC layer on the dielectric leakage current.

The mean values of the leakage current corresponding to Fig. 4 are recorded and presented in Fig. 5, versus the measured CA and the applied voltage, $V$. The CA dependence on the applied voltage for both coatings follows a typical Lippmann behavior up to the critical voltage, $V_s$=45 V, of CA modulation range of 35°. One can observe that the mean leakage current values for the composite coating are slightly higher compared to the Teflon® case. Our measurements show that saturation is followed by a significant leakage current increase. Up to $V_s$, the measured equilibrium current is lower than 20 nA. Beyond the critical applied voltage, where the measured CA does not further decrease, i.e. approximately $1.4V_s$, the current increases by almost an order of magnitude up to 200 nA.

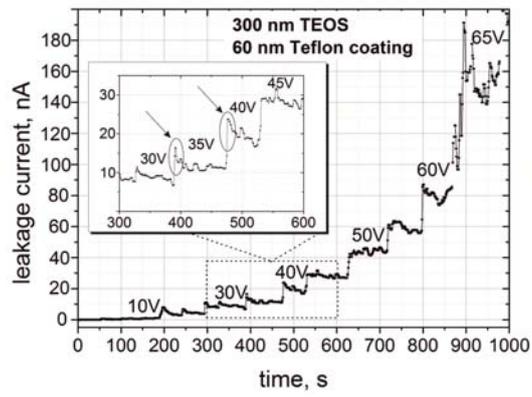

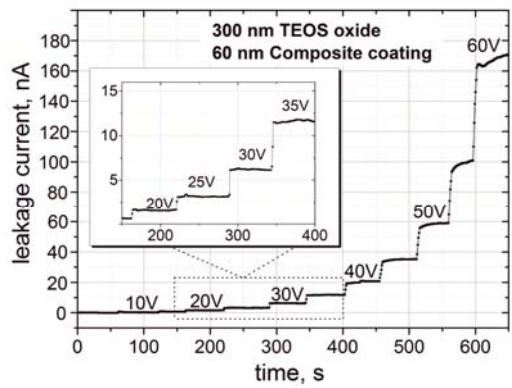

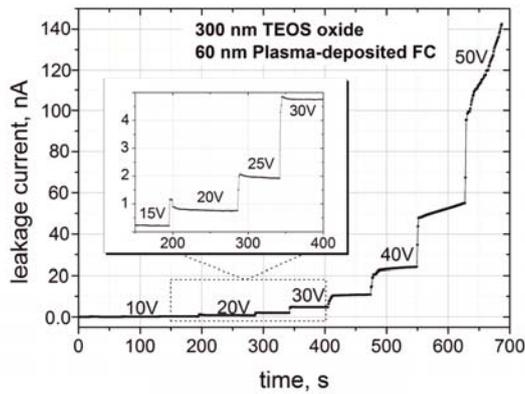

**Figure 4.** Leakage current measurements on stepwise applied voltage increments for the (a) Teflon®, (b) composite and (c) plasma-deposited FC coatings.

For applied voltage higher than $1.4V_s$, there is a vast increase of the current, which is also indirectly detected by bubbles rising in the vicinity of the contact line of the liquid drop. However this indication of electrolysis is only evident at the Teflon® coating, as discussed previously in the text.

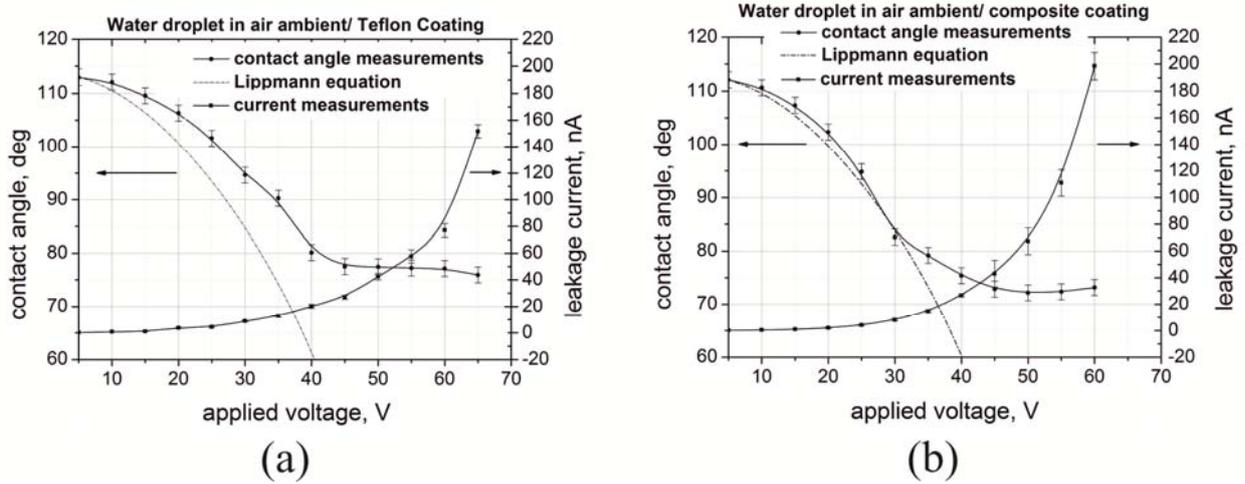

**Figure 5.** Contact angle (left axis), $\theta_V$, and leakage current (right axis) dependence on DC electrowetting on 300nm TEOS featuring (a) Teflon® and (b) composite coatings. (Solid lines are drawn as guide to the eye.)

In order to further demonstrate the superiority of the composite coating compared to the Teflon® coating in terms of dielectric breakdown and sample robustness, we conduct the following test. A sequence of three consecutive DC electrowetting cycles from 0 V to 65 V is performed. During each cycle, we measure the leakage current. In Fig. 6 the normalized leakage current density is shown for each EW cycle. To compute current densities, the measured current is divided by the liquid/solid interfacial area. For each voltage step the current density is normalized by dividing with its value as measured during the first EW cycle.

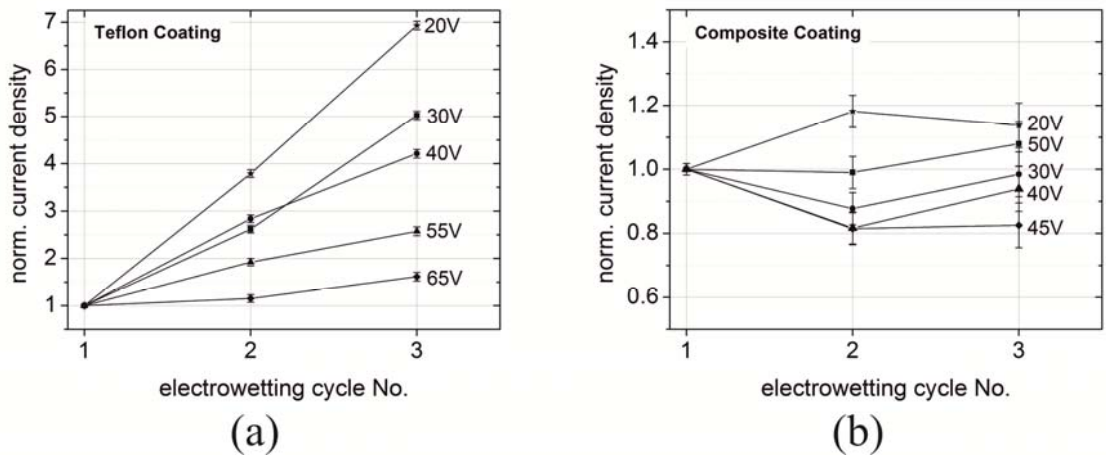

**Figure 6.** Normalized leakage current density as measured during each of the three subsequent DC electrowetting cycles for (a) Teflon[®] and (b) composite coatings.

The repeatability of the measured leakage current densities in a sequence of EW cycles can be used as an indication of dielectric robustness. The leakage current density of the Teflon[®] coating increases (see Fig. 6a), whereas for the composite coating case it remains fairly constant (see Fig. 6b, notice the difference in the limits of y axes of the normalized current density for Teflon[®] and composite coatings between Fig. 6a and 6b). The increasing current density of the Teflon[®] coating suggests wearing out of the dielectric. On the contrary, constant current density of the composite coating signifies preservation of the dielectric properties i.e. improved sample robustness. More specifically, in the case of the Teflon[®] coated sample, the leakage current density during the 3$^{rd}$ cycle increases by up to 7 times its value measured at 1$^{st}$ cycle (at 20 V of applied voltage). It is worth noting that the composite coating showed systematically higher leakage currents than the Teflon coating (see Fig. 5). However, not only the saturation[20] but also the dielectric failure caused by subsequent electrowetting tests, presented in this manuscript, showed remarkable relative increase of the leakage currents. Therefore what mostly matters is the relative increase of the leakage current and not its absolute value which probably depends on the specific material tested. Again, the significantly improved performance in terms of stability of the composite coating can be attributed to its improved adhesion on the TEOS substrate, presumably as a result of the chemical bonding between the underlayer (TEOS) and the plasma-deposited fluorocarbon[15].

Usually in EW devices, relatively thick top coatings are used (>500nm) to achieve the desired EW robustness reversibly. However, the use of the composite coating would suggest thinner top coatings with equally robust performance in consecutive EW cycles, due to consistent leakage current density performance. Thus, sufficiently lower applied voltages would be required. Consequently, the use of the composite coating could be advantageous in low voltage EW systems and devices, featuring improved endurance at multiple EW cycles.

**C. Electrowetting lifetime tests**

Lifetime EW tests are performed in dodecane ambient by continuous long term application (tens of minutes or even hours) of DC or AC voltage. The EW tests were performed for a sessile drop of 1.00wt.% SDS in 0.1M NaCl, on 300nm TEOS/ 60nm composite coating dielectric stack.

We perform these tests in order to present results that are directly comparable with those presented in Dhindsa et al.[14] for a 300nm Parylene HT/ 50nm Fluoropel stack. Three voltage schemes were investigated: +25V DC, -25V DC, and 25V AC (sine wave 2.3kHz). For positive DC voltage (Fig. 7a) the CA remains almost unchanged at ~63° for 1 min. When the voltage is switched off the drop retracts back to ~144°. Further application of DC voltage results in a CA of 63° without any indication of sample failure prior to 15 min; then significant electrolysis is observed. The application of positive DC voltage is the most intensive test on our dielectric stack, since electrolysis is not observed at 25 V AC or -25 V DC even after 6 h.

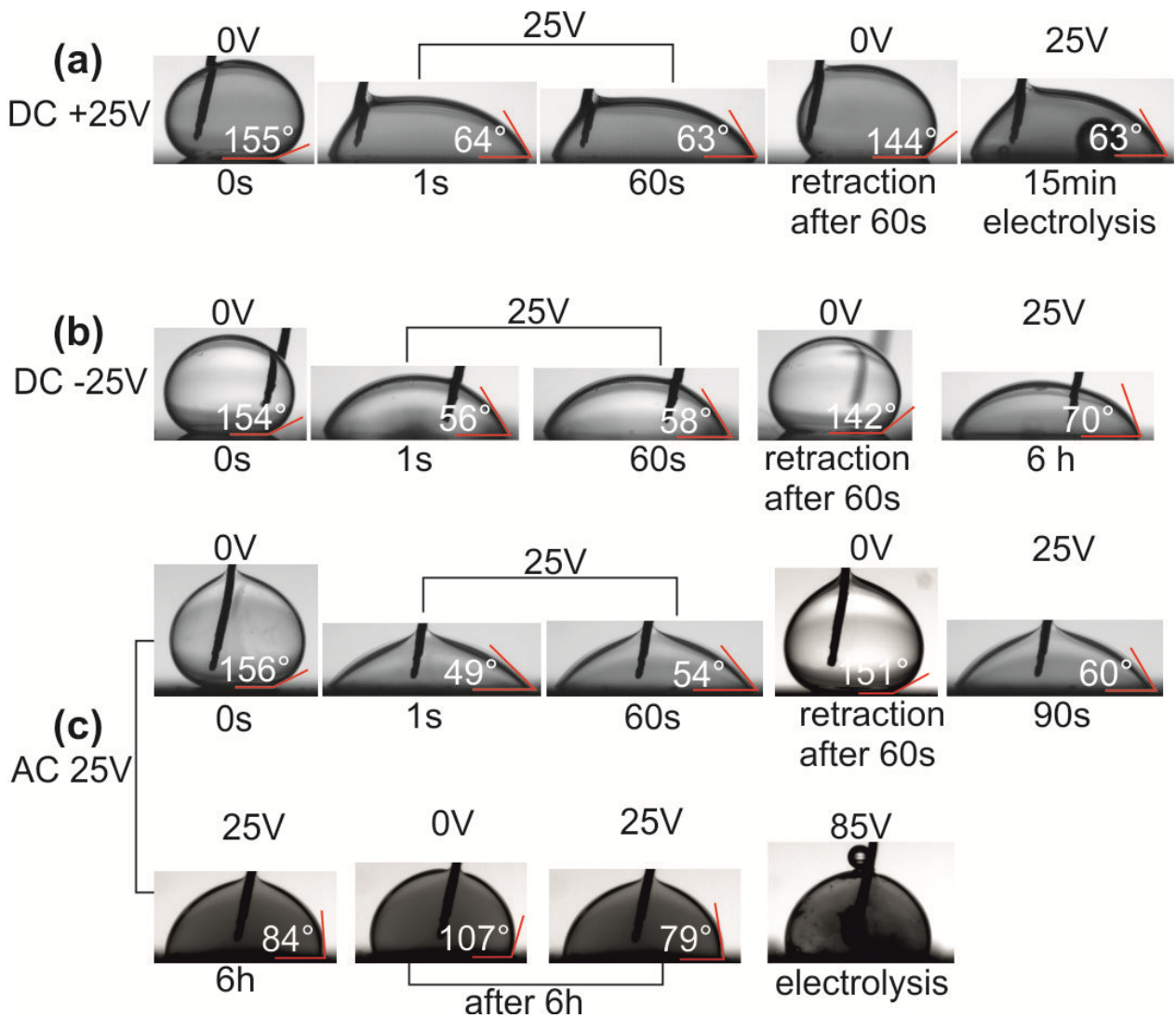

**Figure 7.** Electrowetting lifetime tests for (a) +25 V DC, (b) -25 V DC, and (c) 25 V AC respectively. (averaged CA values are presented over three series of experiments).

For negative DC voltage (Fig. 7b) the sample is capable of exhibiting a CA of ~58° for at least 1 min. By switching the voltage off the drop retracts back to ~142°. Further voltage application results in a CA of 70° which remains unchanged for 6 h; upon voltage removal the drop retracts back to ~125° (image not shown in Fig. 7b). Therefore, since electrolysis is not evident, the sample is mostly reliable with CA modulation range ~55° even after 6h of continuous voltage application. These results are in agreement with the ones presented in Dhindsa et al. where it is shown that negative DC is connected to lower EW hysteresis and higher lifetime possible attributed to less charge trapping.

For AC voltage (Fig. 7c) the sample exhibits a CA of ~49° which gradually increases up to ~54° at 60 s and when the voltage is switched off the drop retracts back to ~151°, revealing low EW hysteresis. Even after 6 h no electrolysis is observed while the sample exhibits a CA of ~84°. Upon voltage removal the drop retracts back to ~107°. Then five EW cycles from 0 V to 25 V were performed with consistent CA modulation range of ~28°. After the completion of the EW cycles, we intentionally increased the voltage up to 85 V, causing electrolysis.

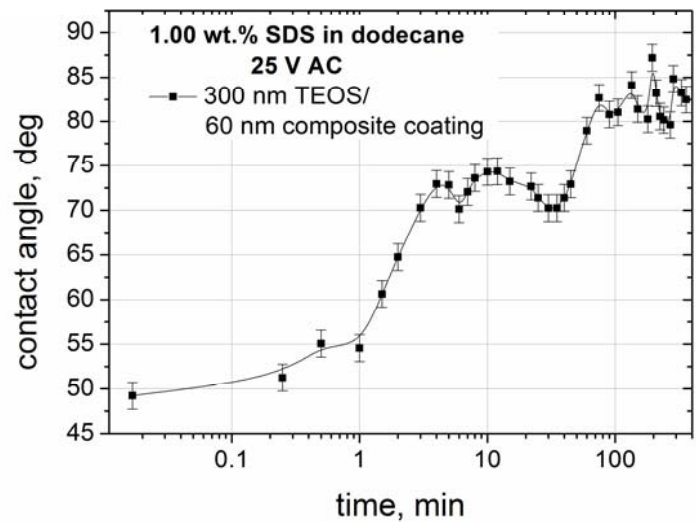

**Figure 8.** Evolution of contact angle in AC electrowetting lifetime tests. (the x-coordinate is in log scale)

Interestingly enough, in the case of continuous AC voltage application, the CA gradually increases (in agreement with the behavior described in Fig. 3) from ~49° to ~72° during the first 30 min and to ~84° after approximately 1.5 h (see Fig. 8). Then the CA remains almost constant until the end of the test. This behavior, possibly attributed to charge trapping, is not observed in the case of -25V DC test where a CA of 70° remains constant for up to 6 h. As mentioned above the sample did not undergo electrolysis at 25 V AC in contrast to the results reported in Dhindsa et al. in which dielectric failure is apparent after 2 h. Hence, the proposed composite coating performs well at both negative DC and AC electrowetting evidently suggesting its implementation on low voltage EW systems.

## Conclusions

We have demonstrated improved robustness of thin hydrophobic dielectric stacks featuring plasma-deposited fluorocarbons on electrowetting tests. The proposed hydrophobic top coating -here called composite coating- demonstrates remarkable electrowetting performance significantly improved compared to spin-coated Teflon® in terms of resistance to dielectric breakdown at voltages beyond the contact angle saturation onset. High consistency of the measured leakage currents after consecutive electrowetting tests indicates enhanced sustainability of the dielectric properties of the composite coating. The improved performance of the composite coating can be connected with our previous finding of its improved adhesion to the TEOS underlayer, possibly attributed to the chemical bonding of the plasma-deposited film to the underlayer[12]. Endurance electrowetting tests (i.e. multiple electrowetting cycles at high applied voltages) reveal no indication of material failure (i.e. electrolysis) with consistently high electrowetting contact angle modulation range (>100°). Furthermore, electrowetting lifetime tests (i.e. continuous long term application of DC or AC voltage at ~$1.3V_s$) exhibited increased reliability in time. It is believed that integration of the composite coating in devices may be proven beneficial to the overall efficiency and robustness of electrowetting systems since the main issue of the bad adhesion of fluoropolymer coatings to oxide substrates is successfully addressed. The suggested coating is currently tested in microfluidic systems where droplet handling (splitting, joining, and movement) is of primary concern.

## Acknowledgements

The research leading to these results has received funding from the European Research Council under the European Community's Seventh Framework Programme (FP7/2007-2013) / ERC Grant agreement n° [240710].